\documentclass[aps,pre,preprint,showpacs,unsortedaddress,
superscriptaddress]{revtex4-1}
\usepackage{epsfig}
\usepackage{latexsym}
\usepackage{amsmath}
\usepackage{amsfonts}
\usepackage{amssymb}
\usepackage{amsbsy}
\usepackage{subfigure}
\usepackage{bm}
\usepackage{color}
\usepackage{dcolumn}
\usepackage{hyperref}
\begin{document}
\title{Sheared phase-separating binary mixtures with surface diffusion}
\author{G. Gonnella}
 \affiliation{Dipartimento di
Fisica, Universit\`{a} di Bari,
 {\it and} INFN, Sezione di Bari,
Via Amendola 173, 70126 Bari, Italy}
\author{A. Lamura}
\email[]{Corresponding author: antonio.lamura@cnr.it}
\affiliation{
Istituto Applicazioni Calcolo, CNR,
Via Amendola 122/D, 70126 Bari, Italy}
\date{\today}
\begin{abstract}
The phase-separation process of a binary mixture with order-parameter-dependent
mobility under shear flow is numerically studied. The ordering is characterized 
by an alternate stretching and bursting of 
domains which produce oscillations in the physical observables.
The amplitude of such modulations reduce in time when the mobility 
vanishes in the bulk phase,
disfavoring the growth of bubbles coming from bursted domains.
We propose two equations for  the typical sizes $R_x$ and
$R_y$ of domains
finding the long-time behaviors
$R_x \sim t^{5/4}$ and $R_y \sim t^{1/4}$ in the flow and shear
directions, respectively, in the case of surface diffusion.
A reduction of the excess viscosity with increasing shear rate is observed
in simulations.
\end{abstract}
\maketitle
\section{Introduction}

The phase separation process when a mixture made of two components
is suddenly quenched below the critical temperature 
is a well studied and understood phenomenon \cite{GUNTON}.
Immediately after the quench, coherent regions rich in either of the phases
form and grow. Typically, if scaling holds,
these domains can be characterized in terms of a single
time-dependent length scale $R(t)$ which grows at late times with a
power law $R(t) \sim t^{\alpha}$, where $t$ is time. 
The problem is very interesting and attracts a lot of interest
\cite{tanaka,lamura08,singh,lamura10,prachi,roy,roy2,jung,tsukada}.

In the case of binary mixtures with constant mobility, 
the value of the exponent $\alpha$
depends on some general properties of the considered system. 
For a nonconserved scalar order 
parameter, as in ferromagnets, it results to be $\alpha=1/2$
\cite{GUNTON}.
When the order parameter is conserved and hydrodynamic effects are ignored,
is found $\alpha=1/3$ \cite{BRAY}. 
External shear flow is known to modify deeply
phase separation producing elongated domains as 
in simulations \cite{simul1,simul2}
and experiments \cite{exp,han,derks}. Numerical solutions of the 
scalar order parameter dynamics
\cite{noisim} as well as analytical \cite{rapa} and numerical 
calculations \cite{noilarge} in the large-$N$ limit of an $N$-vector
order parameter, which corresponds to the
self-consistent approximation of the scalar case,
gave confirmation of the existence of anisotropy in the growing domains
with two characteristic lengths and two different exponents.
The growth exponent is not affected in the shear direction perpendicular
to the flow while in the large-$N$ limit the exponent 
along the flow direction is increased by $1$ 
\cite{rapa,noilarge}. Moreover, all the
physical observables are modulated by oscillations on a logarithmic time
scale which are related to a cyclic mechanism of storage and dissipation
of elastic energy.
 
A more realistic
phase separation scenario, that
takes into account 
the diffusion of molecules of different species
along interfaces between domains of different composition, 
can be obtained by considering an 
order-parameter-dependent mobility 
$\Gamma(\varphi) \propto [1 - c(T) \varphi^2]$ \cite{langer} 
where $\varphi$ is 
the difference of concentrations of the two components of the mixture.
The term $c(T) \rightarrow 1$ for temperature $T \rightarrow 0$
and  $c(T) \rightarrow 0$ for $T \rightarrow T_c$, $T_c$
being the critical temperature. 
This form of the mobility
is such that in the case of a deep quench ($T \rightarrow 0$) 
the diffusion in the bulk, where $\varphi \simeq 1$,
is suppressed while it is favored
along interfaces where $\varphi \simeq 0$. 
This latter mechanism, known as surface 
diffusion (SD), has growth exponent $\alpha=1/4$ \cite{furukawa}
and has relevant technological and scientific applications, for example, 
in deep quenches of polymer mixtures \cite{wolterink}
and coarsening of porous structures by surface diffusion
\cite{geslin}.
On the other hand, for shallow quenches ($T \lesssim T_c$) 
the mobility stays constant and phase separation proceeds
by the Lifshitz-Slyozov mechanism
\cite{lifshitz} with the aforementioned growth exponent $\alpha=1/3$. 
In this case the growth of domains
is related to bulk diffusion (BD), since molecules 
of one component evaporate from
more curved interfaces and diffuse through the bulk of the other 
component diminishing the curvature of the interface.
We add for completeness that in the case of asymmetric mixtures the minority
phase forms isolated droplets. These collide due to Brownian motion and 
coalesce with the same growth exponent of the Lifshitz-Slyozov mechanism
\cite{binder}. 
Numerical studies of phase separation in binary mixtures
with field-dependent mobility
found confirmation of the exponent
$\alpha=1/4$ for $c=1$ as expected for a scalar order parameter
\cite{lacasta,puri,zhu,cast1,gemmert,wolterink}. 
For intermediate values $0 < c < 1$ 
there is a crossover between $\alpha=1/4$ and $\alpha=1/3$.
In the case of an $N$-dimensional order parameter with mobility
given by $\Gamma({\bm \varphi})=1-c {\bm \varphi}^2/N$ simulations
gave $\alpha=1/6$ for $c=1$ and a crossover
from $\alpha=1/6$ to $\alpha=1/4$ for $0 < c < 1$ with $N=2, 3, 4$
\cite{corberi}. We remark that the diffusive exponent of a vectorial order 
parameter is also $1/4$, due to a different physical mechanism 
form that of SD.
A more general form 
$\Gamma({\bm \varphi})=[1-{\bm \varphi}^2/N]^{\beta}$ with 
$\beta > 0$ has been
also introduced and solved in the scalar case $N=1$ \cite{emmott1}
and in the large-$N$ limit \cite{emmott2}.
Effects of shear flow have been considered
in the large-$N$ limit for this latter form of the mobility
that includes the standard case with $\beta=1$. 
It is found that domains grow with 
characteristic length $t^{(2\beta+5)/2(\beta+2)}$ along the flow
direction and $t^{1/2(\beta+2)}$ along the shear
direction with logarithmic corrections
\cite{rapapa} and oscillations of the physical observables
appear to be damped \cite{largen}.
However, domain morphology during phase separation
cannot by fully outlined when using the large-$N$ limit
 since interfaces are missed in the vectorial model 
\cite{BRAY}. For this reason we want to consider the effects of shear flow
on the phase separation process in the case of a scalar order parameter
with field-dependent mobility by solving numerically
the phenomenological equation of the order parameter in two spatial
dimensions. The novelty of the present study relies on addressing
for the first time the effects of surface diffusion
in the simulation of a phase separating 
binary mixture under shear.

The paper is organized as follows. Section II describes the model and 
Sect.~III is devoted to present and discuss
numerical results.
Phenomenological equations for the typical
sizes of domains suggest
domain growth with power laws $t^{5/4}$ and 
$t^{1/4}$ in the flow and shear
directions, respectively.
Simulations give compatible results and
provide direct evidence
of a cyclic stretching and bursting of 
domains which are responsible of oscillations in the physical observables.
The amplitude reduces in time due to the vanishing 
of the mobility in the bulk phase
which makes difficult the growth of bubbles coming from bursted domains
thus finding confirmation of the argument put forward in Ref.~\cite{largen}.
Finally, we draw conclusions.

\section{The model}

The equilibrium properties of the system we are considering,
are described by the
Ginzburg-Landau free-energy
\begin{equation}
{\cal F}\{\varphi\} = \int d {\bm r}
\left \{- \frac{a}{2} \varphi^2 + \frac{b}{4} \varphi^4
+ \frac{\kappa}{2} \mid \nabla \varphi \mid^2 \right \} 
\label{eqn1}
\end{equation}
where $\varphi$ is the order parameter. 
As usual, it is assumed $b>0$ ensuring stability.
The coefficient $a=(T_c-T)/T_c$ can be considered as a reduced temperature
where $T$ is the temperature of the system and
$T_c$ is the critical value below which the fluid is ordered.
When $T<T_c$ the polynomial terms of the free-energy density have
two symmetric minima located
at $\varphi_{eq}(T)=\pm \sqrt{a/b}$ which are the equilibrium values
of the order parameter. 
Finally, the parameter $\kappa >0$ encodes 
the energy cost for the creation of interfaces between domains
of different composition.
The chemical potential difference between the two components can be
computed from Eq.~(\ref{eqn1}) and is given by
\begin{equation}
\mu =  \frac{\delta{\cal F}}{\delta\varphi}= - a \varphi +
b \varphi^3 - \kappa \nabla^2 \varphi .
\label{mu}
\end{equation}

The time evolution of the order parameter $\varphi$ 
is described by the convection-diffusion
equation
\begin{equation}
\frac {\partial \varphi} {\partial t} + {\bm \nabla} \cdot (\varphi {\bm v}) =
{\bm \nabla} \cdot \Big [ \Gamma(\varphi) 
{\bm \nabla} \mu \Big ] .
\label{eqn2}
\end{equation}
The convective term on the l.h.s. couples $\varphi$ 
to the external velocity field which is given here
by a linear shear flow ${\bm v} = \dot\gamma y {\bm e_x}$ 
where $\dot\gamma$ is the shear rate, $y$ the coordinate along the
$y$-direction (shear direction), and ${\bm e_x}$ the  unit vector along the 
$x$-direction (flow direction).
We assume the fluid to be very viscous so that the 
Reynolds number (ratio of inertial to viscous forces) 
is very low and the capillary number (ratio of viscous to interfacial 
(in two dimensions) forces)
is very large. In this way
hydrodynamic effects are neglected in the present model so that
Eq.~(\ref{eqn2}) is not coupled to the Navier-Stokes equation
for the flow field. 
Very recently the general case, when the capillary number 
is not large, has been considered
in Ref.~\cite{chih} for the phase separation of sheared binary mixtures.
We also neglect thermal fluctuations.
The mobility depends explicitly
on the order parameter in order to have a more realistic
description of the phase separation, especially for polymer mixtures
\cite{langer}, and is given by
\begin{equation}
\Gamma(\varphi)=\Gamma_0 \Big [ 1-\varphi^2/\varphi_{eq}^2(0) \Big ] 
\end{equation}
where $\Gamma_0$ is a constant.
We comment here about the role of the temperature on the mobility
when there is no external flow (${\bm v}=0$). 
In the case of a deep quench ($T \rightarrow 0$), 
the diffusion in the bulk, where $\varphi \simeq \varphi_{eq}$, 
is
suppressed since 
$\Gamma(\varphi_{eq}(T))
 \rightarrow 0$ promoting
surface diffusion. 
On the other hand for shallow quenches ($T \rightarrow T_c$), 
the mobility
is not significantly reduced in the bulk since 
the $\varphi_{eq}^2(T) \ll \varphi_{eq}^2(0)$.
In this limit the Cahn-Hilliard equation \cite{cahn} is recovered whose
numerical solution \cite{rogers} gives evidence of the growth
exponent $\alpha=1/3$ typical of bulk diffusion. 

Equation (\ref{eqn2}) can be written in a dimensionless form \cite{rogers}
after redefining time, space, and field scales by 
$\tau=\kappa/(2 \Gamma_0 a^2)$, $\xi=\sqrt{\kappa/a}$, and 
$\varphi_{eq}(T)$, respectively.
In the following the symbol \^{}
will denote dimensionless quantities. The only relevant parameters
in the dimensionless equation 
are the mobility and the shear rate.
The mobility assumes the form 
$\hat\Gamma(\hat\varphi)=1-\lambda \hat\varphi^2$, where
$\lambda=\varphi_{eq}^2(T)/\varphi_{eq}^2(0)$ goes from $0$ to $1$ 
when the temperature is reduced from $T_c$ to $0$, the bulk
equilibrium values of the order parameter are $\pm 1$, and
the shear rate is $\hat{\dot\gamma}=\dot\gamma \tau$.

\section{Numerical results and discussion}

We have simulated Eq.~(\ref{eqn2}) in two dimensions 
by using a finite-difference scheme.
The field $\varphi({\bm r},t)$ is discretized on the nodes $(x_i,y_j)$
($i,j=1,2,...,L$) of a square lattice with space step $\Delta x$
and $L \times L$ nodes. Time is discretized in time steps $\Delta t$.
Periodic boundary conditions were implemented
in the $x$-direction while Lees-Edwards
boundary conditions \cite{LE} were used in the normal direction. The latter
conditions require the identification of a
point on the lower row at $(x_i,y_1)$ with the one located on the upper row
at $(x_i+\dot\gamma L \Delta t, y_L)$ ($i=1,2,...,L$) 
to take into account the space shift
due to shear. The time derivative in Eq.~(\ref{eqn2}) was implemented by using
an explicit first-order Euler algorithm \cite{rogers} while
standard central difference schemes were adopted for spatial derivatives
appearing in the convective and diffusion terms \cite{strik}.

Simulations were run using lattices with $L=1024$. 
We fix $a=b=\kappa=\Gamma_0=1$
and write the mobility in Eq.~(\ref{eqn2}) as
$\Gamma(\varphi)=1-\lambda \varphi^2$ with $0 \leq \lambda \leq 1$. 
This form of the mobility allows us
to consider
a deep quench ($\lambda=1$)
where SD is the leading diffusion mechanism.
In this case the numerical model can become
unstable when $|\varphi| > 1$, 
due to numerical fluctuations,
since the mobility results to be negative. 
To overcome this problem we used a fine mesh with $\Delta x = 0.5, 1$
and a very small time step $\Delta t=10^{-3}$ enforcing the value $1$ 
($-1$) to $\varphi$ whenever $\varphi > 1$ 
($\varphi < -1$). 
Neither relevant differences were
observed for the two values of $\Delta x$ nor significant violation
of field conservation could be appreciated. 
Equation (\ref{eqn2}) was also solved by using intermediate 
values $\lambda < 1$. The limit case
$\lambda=0$ allows the simulation
of a quench where BD dominates the phase separation process.
In this way it was possible 
to consider the crossover between surface and bulk diffusion mechanisms
under the presence of shear.
The system was prepared in
a disordered state ($T > T_c$) with $\langle \varphi \rangle = 0$
corresponding to a symmetric composition of the mixture, 
$\langle ... \rangle$ denoting an average over the system.
The results here exposed were obtained with $\Delta x=1$, $\dot\gamma$
in the range $[10^{-3},5 \times 10^{-2}]$, $\lambda=0, 0.4, 0.8, 1$, 
and averaging over five
independent realizations of the system. All the quantities in the 
following are expressed
in units of $\Delta x$, $\Delta t$, and $\varphi_{eq}$.

\begin{figure}[ht]
\epsfig{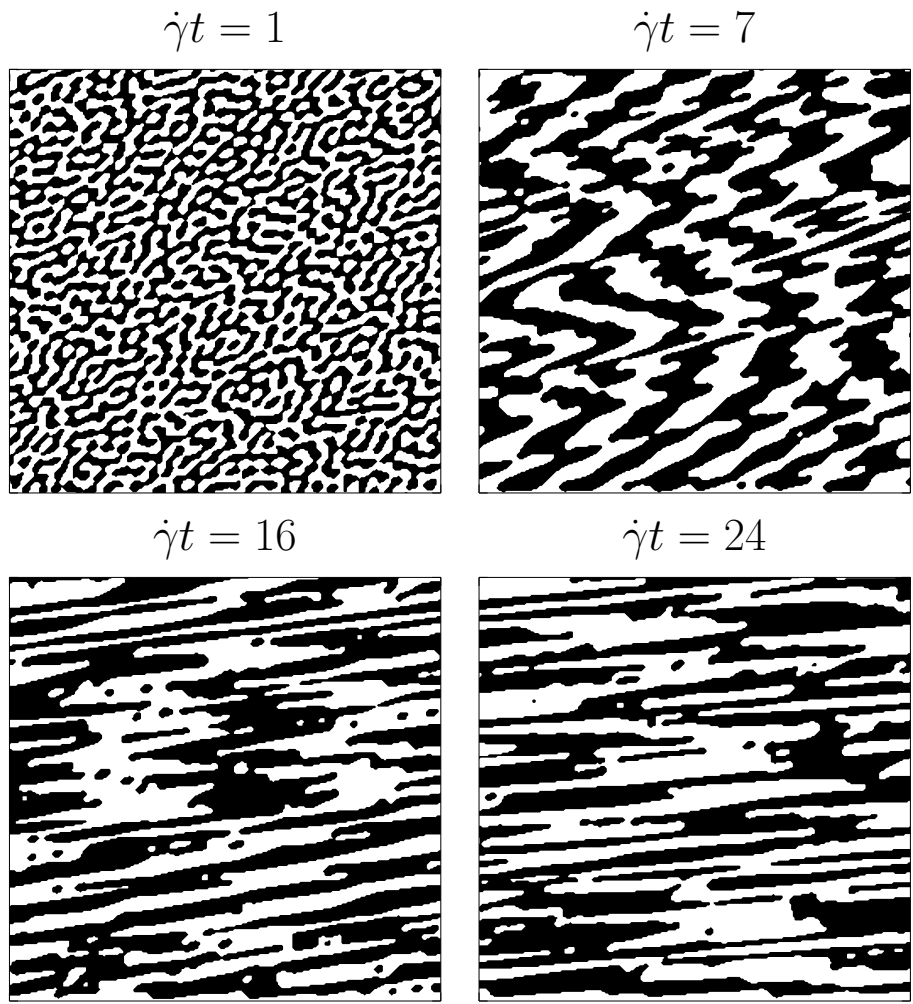}
\caption{Configurations at consecutive times 
of the system in the case with $\lambda=1$. 
Black/white domains correspond to positive/negative values
of the order parameter $\varphi$. A portion of size $256 \times 256$
of the whole lattice is shown.
}
\label{fig1}
\end{figure}

\begin{figure}[ht]
\epsfig{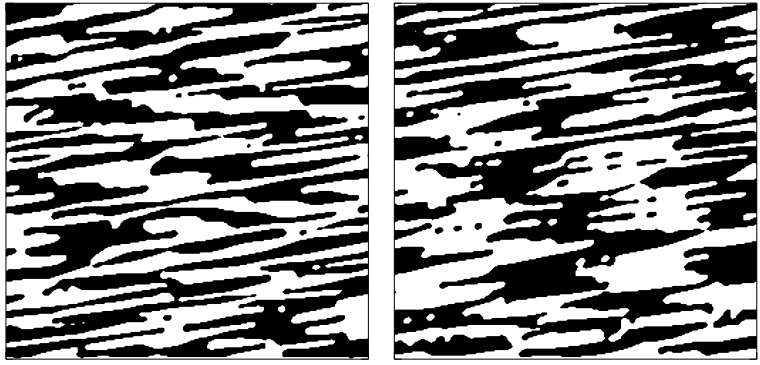}
\caption{Configurations at $\dot\gamma t=16$ 
of the system in the cases with $\lambda=0$ (left) and
$0.8$ (right).
Black/white domains correspond to positive/negative values
of the order parameter $\varphi$. A portion of size $256 \times 256$
of the whole lattice is shown.
}
\label{fig2}
\end{figure}

The morphology of the phase-separation process in the SD case ($\lambda=1$)
is shown for $\dot\gamma=5 \times 10^{-2}$ in Fig.~\ref{fig1}
at consecutive values of the strain $\dot\gamma t$ on a portion
of the whole lattice of size $256 \times 256$.
At the beginning a bicontinuous structure is formed yet with no
appreciable deformation induced by shear. For values $\dot\gamma t > 1$ 
the external flow starts to modify domains which are elongated, tilted, 
and then bursted by shear when accumulated stress overcomes
surface tension. 
At $\dot\gamma t=7$ it can be noted that domains are stretched by the
flow and characterized by different thicknesses. 
At later times the flow further deforms domains which 
may break up,
due to the aforementioned mechanism, giving rise
to several bubbles which
spread all over the system ($\dot\gamma t=16$).
In the SD case the suppression of bulk diffusion
inhibits the growth of these 
small bubbles originating from the shear-induced
bursting of domains. 
As a matter of comparison we plot in Fig.~\ref{fig2}
the configurations at the same value of the strain for the values
$\lambda=0, 0.8$. It is evident that by decreasing $\lambda$, the number
of bubbles in the system reduces being minimum in the case with $\lambda=0$
where bubbles can grow by bulk diffusion. 
This feature will be further discussed in the following.
As a consequence, 
bubbles cannot be further deformed in the SD case
by the flow but can eventually
merge with larger domains when these are stretched 
again by the flow and reach some bubble ($\dot\gamma t=24$).
In the large-$N$ limit \cite{largen} it was found that 
the phenomenon of elongation and bursting
of domains is cyclic.
Here we find an indication of this periodic behavior but the
finite size of the computational domain does not allow
the observation on very long times to find full evidence of periodic behavior
on a logarithmic time scale.

\begin{figure}[ht]
\epsfig{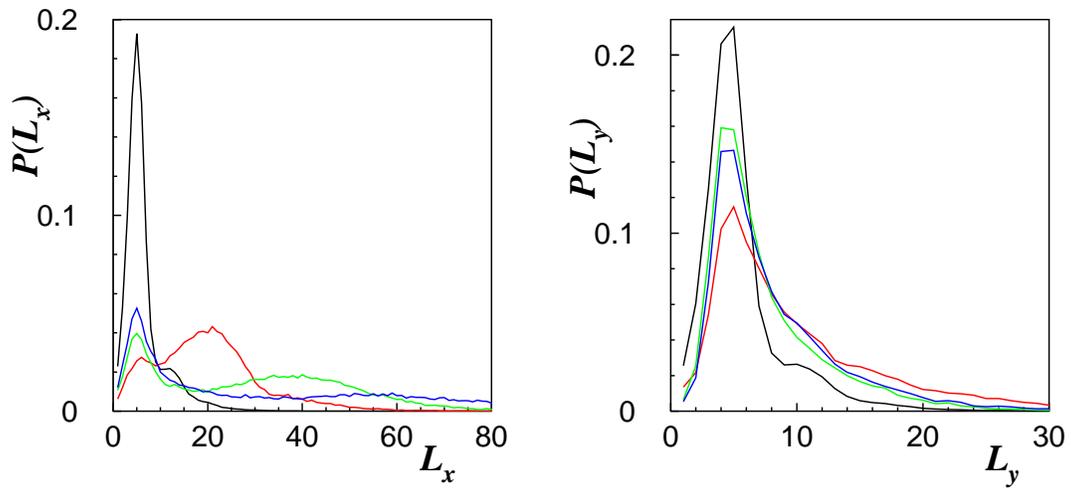}
\caption{Probability distribution functions $P$
of domains
of length $L_x$ (left panel) and $L_y$ (right panel)
at strains $\dot\gamma t= 1$ (black line), $7$ (red line),
$16$ (green line), and $24$ (blue line)
in the case with $\lambda=1$.
}
\label{fig3}
\end{figure}

\begin{figure}[ht]
\epsfig{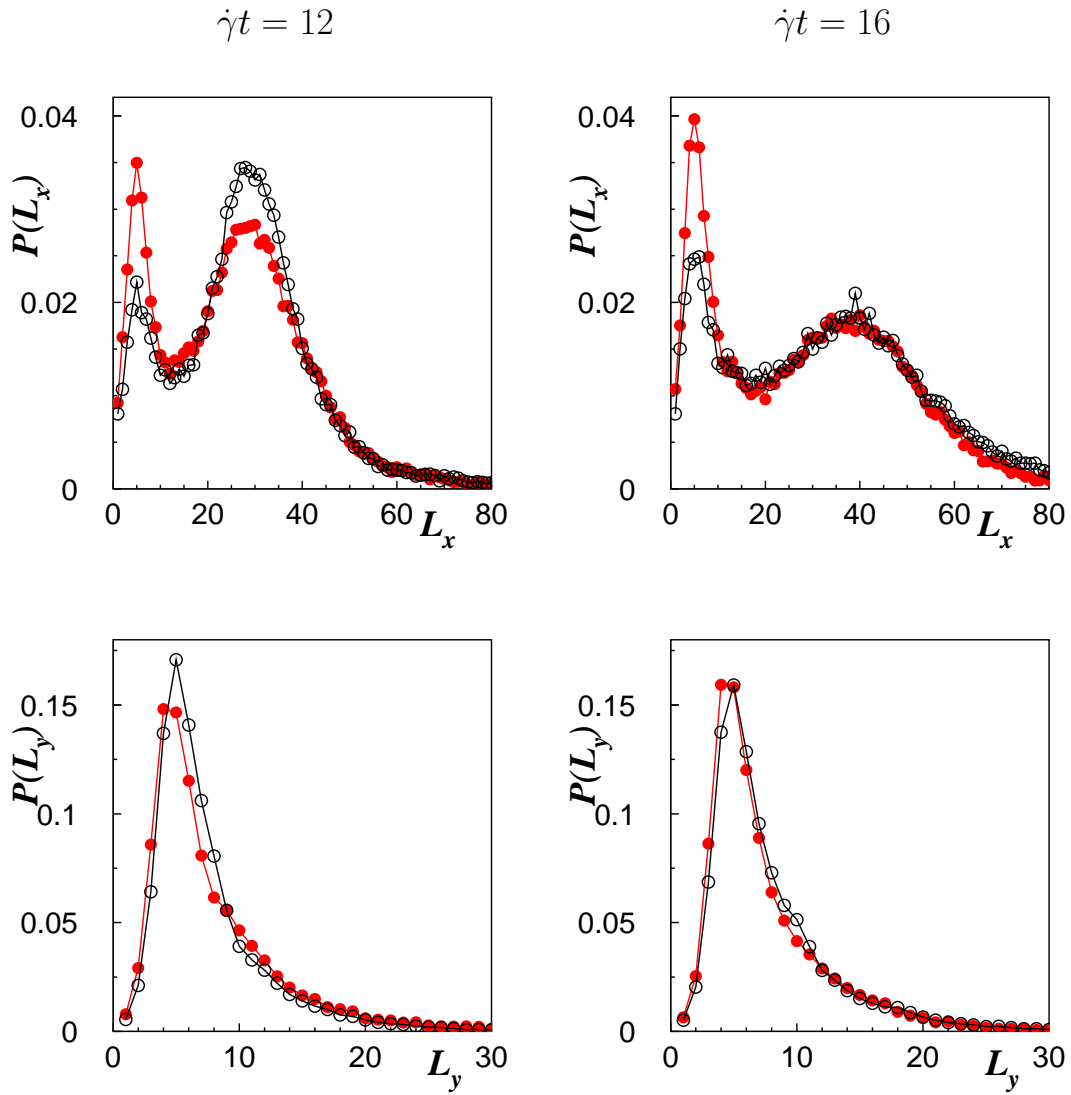}
\caption{Probability distribution functions $P$
of domains
of length $L_x$ (upper panels) and $L_y$ (lower panels)
at strains $\dot\gamma t= 12, 16$
in the cases with $\lambda=0 (\circ), 1 (\bullet)$.
}
\label{fig4}
\end{figure}

In order to characterize the size distribution 
of growing domains, 
the normalized probability distributions functions $P(L_{x,y})$ of finding
domains of lengths $L_x$ and $L_y$
along the flow and the
shear directions, respectively, were computed.
First we estimated the extensions $L_x$ 
of the unidimensional domains of different composition
along all the rows of the computational mesh. 
From the recorded values of $L_x$, we finally computed $P(L_x)$. The 
same procedure was followed along the shear direction to compute $P(L_y)$.
The plots of $P(L_x)$ and $P(L_y)$
are shown in Fig.~\ref{fig3} at different times with $\lambda=1$.
At $\dot\gamma t=1$ both $P(L_x)$ and $P(L_y)$
have a single main peak at $L_{x1} \simeq L_{y1} \simeq 5$, 
corresponding to initial isotropic domains, and a small shoulder in the tail. 
Later anisotropy effects become
evident in the distributions.
Along the flow direction 
it is possible to observe at $\dot\gamma t=7$ 
the formation of a second peak at 
$L_{x2} \simeq 20$,
corresponding to more elongated domains, which prevails on the other one at 
$L_{x1}$.
Along the shear direction, at the same time, the main peak decreases in
height while the distribution broadens corresponding to the presence of domains
with different thicknesses. 
The further stretching of domains is such that the peak
at $L_{x2}$  moves at following time towards a larger
value ($\simeq 40$) while the breaking of elongated
domains, which promotes the formation of bubbles, is such that 
the peak at $L_{x1}$ dominates (see $\dot\gamma t=16$ in Fig.~\ref{fig3}). 
This supports the previously described picture where
one can observe an alternate dominance of elongated and
bursted domains. 
At the same time
the peak of $P(L_y)$ at $L_{y1}$ grows again while the tail decays more rapidly
with $L_y$ than at the previous time
as consequence of the formation of 
bubbles of typical size $L_{x1} \simeq L_{y1}$. At the last time
considered in Fig.~\ref{fig3})
the peak at $L_{x1}$ of $P(L_x)$ still grows while the  second peak moves
at $L_{x2} \simeq 60$, since domains continue to be stretched, decreasing
in height and broadening. The distribution $P(L_y)$ does not change 
significantly. Figure \ref{fig4}
compares the probability distribution functions
in the BD and SD cases. It is found that the peak
at $L_{x1}$ is always higher in the SD case for $\dot\gamma t > 10$ 
while in the BD case there is an alternate dominance of the peaks
at $L_{x1}$ and $L_{x2}$.
This confirms the larger abundance of bubbles which cannot grow
after their formation since the variable
mobility suppresses diffusion in the bulk. 
Therefore bubbles cannot be elongated too much by the
flow and this also reflected in the distribution $P(L_y)$. Indeed its peak at
$L_{y1}$ does not change significantly in height for $\dot\gamma t > 10$
while in the BD case the peak shows weak oscillations
since domains are continuously stretched
and bursted while growing.

\begin{figure}[ht]
\epsfig{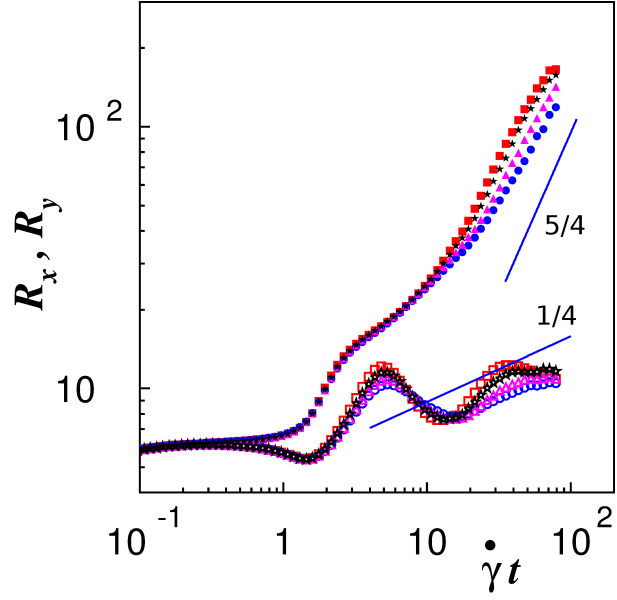}
\caption{Average sizes of domains
along the flow (filled symbols) and the shear (empty symbols)
directions as functions of time
for the cases with $\lambda=0 (\blacksquare), 0.4 (\star), 0.8
(\blacktriangle), 1 (\bullet)$.
The lines have slopes $1/4$ and $5/4$.
}
\label{fig5}
\end{figure}

In the isotropic case without shear
it was proposed the following equation
for the domain size $R(t)$ \cite{lacasta}:
\begin{equation}
\frac{d R(t)}{d t}=
(1-\lambda) \frac{A}{R^2}+\lambda \frac{B}{R^3}
\label{r}
\end{equation}
where the first and second terms on the r. h. s. take into account
the bulk and surface diffusion, respectively, $A$ and $B$ being two constants.
It comes out that $R(t)$ grows in time with a power law with exponents
$\alpha=1/4$ for $\lambda=1$ and $\alpha=1/3$ for $\lambda=0$.
A crossover between these two regimes was found in numerical simulations
for intermediate values of $\lambda$ \cite{lacasta}.
In the case with shear, 
denoting by  $R_x$ and $R_y$ the sizes of an elongated domain 
along the flow and the shear directions, 
respectively, 
Eq.~(\ref{r}) can be generalized to:
\begin{eqnarray}
\frac{d R_x(t)}{d t}&=&C_0 \dot\gamma R_y 
+ (1-\lambda) C_1 \Big ( \frac{1}{R_x^2} +\frac{1}{ R_y^2} \Big )+ \lambda
C_2 \Big ( \frac{1}{R_x^3} +\frac{1}{ R_y^3} \Big )  \label{rx}\\
\frac{d R_y(t)}{d t}&=&
(1-\lambda) C_1 \Big ( \frac{1}{R_x^2} +\frac{1}{ R_y^2} \Big )+ \lambda
C_2 \Big ( \frac{1}{R_x^3} +\frac{1}{ R_y^3} \Big )
\label{ry}
\end{eqnarray}
where $C_0$, $C_1$, and $C_2$ are constants.
The first term on the r.h.s. of Eq.~(\ref{rx}) takes into account
the growth along the $x$-direction
caused by the advection of domains due to the shear and is
related to the amount of flow intercepted by the domain which is proportional
to $\dot\gamma R_y$. 
Solving the equations under the hypothesis $R_x \gg R_y$ at long times, 
we find the asymptotic behaviors
$R_x \sim \dot\gamma t^{5/4}$ and $R_y \sim t^{1/4}$ with $\lambda=1$
and recover the results
$R_x \sim \dot\gamma t^{4/3}$ 
and $R_y \sim t^{1/3}$ with $\lambda=0$ \cite{noisim}.
The average sizes $R_x$ and $R_y$ of domains
were computed in simulations in two different ways.
By measuring the total 
lengths $I_x$ and $I_y$ of the interfaces between domains  
along the flow and the shear directions, respectively, we defined
$R_{x,y}=L^2/I_{x,y}$. The second procedure relies on the use of
the structure
factor $C({\bm k},t)= \langle \varphi({\bm k}, t)\varphi(-{\bm k},
t)\rangle $, where $\varphi({\bm k}, t)$ is the Fourier transform
of the order parameter. From this quantity,
the average sizes of domains were computed
as
\begin{equation}
R'_{x,y}(t) = \pi \frac{ \int d{\bm k} C({\bm k},t)}{\int d{\bm k} |k_{x,y}|
C({\bm k},t)} .
\label{eqnrad}
\end{equation}
The time behavior of $R_x$ and $R_y$ is reported in Fig.~\ref{fig5}
for $\lambda=0, 0.4, 0.8, 1$.
We found that $R'_x$ and $R'_y$ have time dependencies
similar to the ones of $R_x$ and $R_y$ 
being $R'_{x,y}(t) \simeq R_{x,y}(t)/3$ over the explored time range.
It can be seen that $R_x \simeq R_y$ until $\dot\gamma t \simeq 1$, 
independently on $\lambda$,
while later
$R_x$ grows faster than $R_y$.
We find that the growth of $R_x$ at long times is slower when increasing
$\lambda$ as expected from the asymptotic solution of Eq.~(\ref{rx}). However,
the limited size of the simulated system does not allow us
to have a reliable estimate of the growth exponent along the flow direction.
The size $R_y$ along the shear direction shows
oscillations which are related to the alternate
stretching (minimum of $R_y$) and breaking (maximum of $R_y$)
of domains. The amplitude of such oscillations shrinks when increasing
$\lambda$.
In the case of surface diffusion ($\lambda=1$) 
oscillations 
appears to reduce significantly in time.
The same behavior was observed in the 
large-$N$ limit \cite{largen} where 
all the physical observables showed 
damped oscillations on a logarithmic time-scale. 
In that model it was guessed that the damping of oscillations
might have been related to the suppression of bulk diffusion. 
In this paper, we directly show that bubbles, 
coming from the bursting of overstretched domains, cannot be
elongated too much by the flow. 
As a consequence, we observe less wide oscillations which
further reduce in time since the mechanism of growth inhibition 
becomes more effective.
Indeed we find evidence, as illustrated in Figs.~\ref{fig1} and \ref{fig3},
of this picture. However, the period of observation, due to the finite
extension of the simulated system, is limited to a bit more than one cycle
of stretching and bursting.

\begin{figure}[ht]
\epsfig{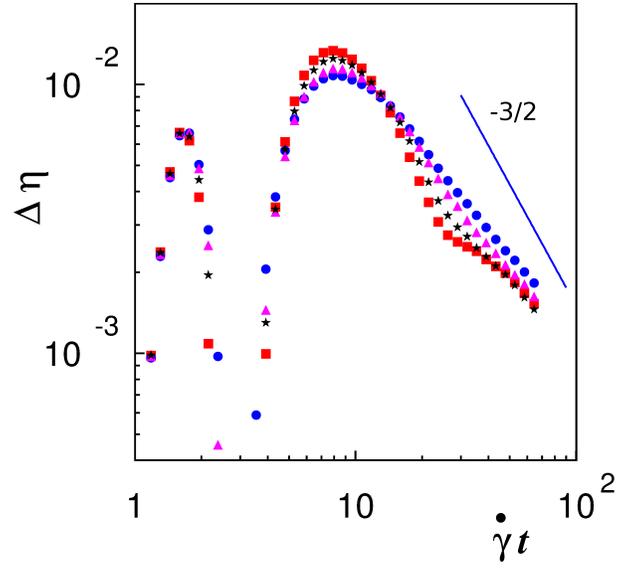}
\caption{Excess viscosity as a function of time
for the cases with $\lambda=0 (\blacksquare), 0.4 (\star), 0.8
(\blacktriangle), 1 (\bullet)$.
The line has slope $-3/2$. 
}
\label{fig6}
\end{figure}

Of experimental interest is the excess
viscosity which measures the variation of the mixture viscosity with
respect to the homogeneous case.
It is defined as \cite{onuki97}
\begin{equation}
\Delta \eta = -\frac{1}{\dot\gamma}
\int \frac{d{\bm k}}{(2 \pi)^2} k_x k_y C({\bm k},t)
\label{eqn5bb}
\end{equation}
whose time behavior is shown in Fig.~\ref{fig6} for different values of
$\lambda$.
We find that the excess viscosity grows oscillating reaching a global
maximum at $\dot\gamma \simeq 8$ when domains are stretched. 
Later $\Delta \eta$
decays to zero due to the dissipation of the energy stored
by elongated domains when they start to burst after further stretching.
It appears that the excess viscosity shows, when increasing $\lambda$,
reduced oscillations in the long time limit, due to
the inhibition of the growth of circular domains
coming from bursting, and a slower time decay.
In the SD case the behavior of the excess viscosity
in the long time limit can be obtained by
assuming that scaling is verified asymptotically. 
This implies that
the structure factor can be written in the anisotropic case as 
$C({\bm k},t) \sim R_x R_y f(k_x R_x,k_y R_y)$ \cite{noisim}.
Then it can be shown that
the excess viscosity $\Delta \eta \sim 1/(\dot\gamma R_x R_y)$
has asymptotic behavior
$\Delta \eta \sim \dot\gamma^{-1/2} (\dot\gamma t)^{-3/2}$ where
the exponent $-3/2$ is, indeed, smaller than the value $-5/3$ valid in
the BD case \cite{noisim}. However, present simulations do not
allow a reliable estimate of the decay exponent.

\begin{figure}[ht]
\epsfig{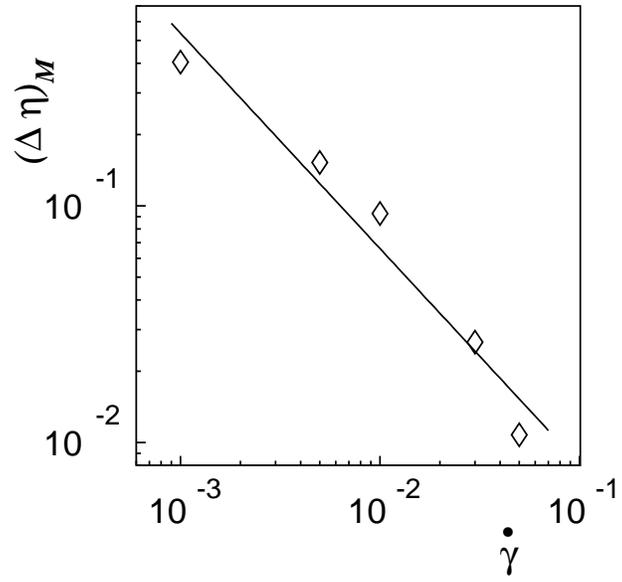}
\caption{The maximum of the excess viscosity as a function
of the shear rate in the case with $\lambda=1$. 
The full line is the best fit
with slope $-0.91 \pm 0.11$.
}
\label{fig7}
\end{figure}

Finally, the values $(\Delta \eta)_M$ of the excess viscosity at 
its maximum are plotted as a function of the shear rate 
in Fig.~\ref{fig7} in the SD case. A reduction of $(\Delta \eta)_M$
can be seen and can be understood in the following way.
The previous scaling analysis of the excess viscosity behavior suggests
that, for a fixed value of the strain $\dot\gamma t$,
the scaling $\Delta \eta \sim \dot\gamma^{-1/2}$ should hold.
We find in simulations an effective exponent $-0.9$ which is smaller than the
expected one. 

\section{Conclusions}

In this paper we have studied numerically the phase separation of a binary
mixture under shear flow when the mobility
depends explicitly on the order parameter. 
The morphology of patterns has been analyzed and discussed
by computing the probability distribution functions of the size
of domains along the two spatial directions.
When surface diffusion dominates the coarsening of the mixture,
it is found that domains are elongated, tilted, and finally 
bursted by the flow.
This process repeats in time producing oscillations 
in physical observables as in the case with constant mobility where
bulk diffusion is the leading mechanism. 
Here the novelty is that bubbles, coming from the disruption 
of stretched domains, cannot grow due to the reduction of the diffusion in
the bulk phase when surface diffusion prevails. As a consequence, 
the oscillation amplitudes in
the size of domains as well as in the excess viscosity
reduce in time going from the bulk to the surface diffusion. This gives
a direct evidence of
the picture put forward in the large-$N$ limit \cite{largen}. 
A generalization of a phenomenological equation
for the typical size of domains has been proposed.
This allows us to obtain
the asymptotic behaviors of characteristic measures of growing patterns
along the flow and the shear directions.
Numerical simulations show that the growth slows down when going from bulk
diffusion to surface diffusion. The same behavior is also observed
for the excess viscosity whose maximum value, at fixed strain,
decreases with increasing shear rate.


\newpage
\clearpage

\newpage
\clearpage

\newpage
\clearpage

\newpage
\clearpage

\newpage
\clearpage

\newpage

\newpage
\clearpage

\end{document}